\documentclass[pre,twocolumn,groupedaddress,showpacs,showkeys,amsmath]{revtex4}

\usepackage{graphicx}
\usepackage{dcolumn}
\usepackage{bm}

\begin{document}


\title{Percolation thresholds on planar Euclidean relative neighborhood graphs}

\author{O. Melchert}
\email{oliver.melchert@uni-oldenburg.de}
\affiliation{
Institut f\"ur Physik, Carl von Ossietzky Universit\"at Oldenburg, D-26111 Oldenburg, Germany\\
}

\date{\today}


\begin{abstract}
In the presented article, statistical properties regarding the topology
and standard percolation on relative 
neighborhood graphs (RNGs) for planar sets of points, considering the Euclidean 
metric, are put under scrutiny.
RNGs belong to the family of ``proximity graphs'', i.e.\ their edge-set encodes
proximity information regarding the close neighbors for the terminal nodes of a given edge.
Therefore they are, e.g., discussed in the context of the construction of backbones for
wireless ad-hoc networks that guarantee connectedness of all underlying nodes.

Here, by means of numerical simulations,
we determine the asymptotic degree and diameter of RNGs and we estimate
their bond and site percolation thresholds, which were previously conjectured to 
be nontrivial. We compare the results to regular $2D$ graphs
for which the degree is close to that of the RNG. 
Finally, we deduce the common percolation critical exponents from the RNG data to 
verify that the associated universality class is that of standard $2D$ percolation.
\end{abstract} 

\maketitle

\section{Introduction \label{sect:introduction}}

The pivotal question in standard percolation \cite{stauffer1979,stauffer1994} is that 
of connectivity. A basic example is $2D$ random bond percolation, where 
one studies a lattice in which a random fraction $p$ of the 
edges is ``occupied''.
Clusters composed of adjacent sites joined by occupied edges are then analyzed 
regarding their geometric properties. 
Depending on the fraction $p$ of occupied edges, the geometric 
properties of the clusters change, leading from a phase 
with rather small and disconnected clusters to a phase, where there is 
basically one large cluster covering the lattice.  
Therein, the appearance of an infinite, i.e.\ percolating, cluster is 
described by a second-order phase transition.  

There is a wealth of literature on a multitude of variants on the
above basic percolation problem that model all kinds of phenomena,
ranging from simple configurational statistics \cite{Ziff2000} to ``string''-bearing 
models that also involve a high degree of optimization, e.g.\ describing
vortices in high $T_c$ superconductivity \cite{pfeiffer2002,pfeiffer2003},
the negative-weight percolation problem \cite{melchert2008,melchert2010a},
and domain wall excitations in disordered media such as $2D$ spin glasses \cite{cieplak1994,melchert2007} and 
the $2D$ solid-on-solid model \cite{schwarz2009}. 
Besides discrete lattice models there is also interest in studying 
continuum percolation models, where recent studies reported on highly
precise estimates of critical properties for spatially
extended, randomly oriented and possibly overlapping objects with various 
shapes \cite{MertensMoore2012}.
Further, percolation phenomena on planar random graphs and their duals
have been studied extensively in the past \cite{Hsu1999,Becker2009,Kownacki2008}. 
Among the latter graphs are, e.g., $2D$ Voronoi graphs related to a planar 
set of points, and their duals, given by the Delaunay triangulation of an associated 
auxiliary set of points \cite{Becker2009}.
In the latter reference the general interest of computing percolation thresholds for 
other random systems is declared.

\begin{figure}[t!]
\centerline{
\includegraphics[width=1.0\linewidth]{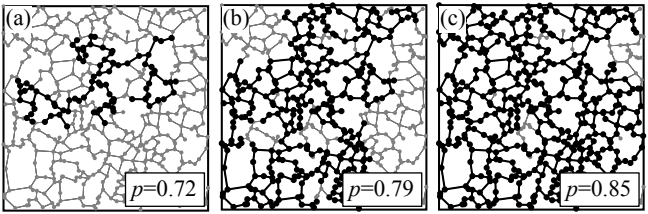}}
\caption{
Bond percolation on an instance of a relative 
neighborhood graph for a planar set of $N=512$ points.
The subfigures relate to a fraction $p$ of occupied edges, where
(a) $p=0.72$ (sub-critical), (b) $p=0.79$ (close to the critical point), and (c) $p=0.85$ (super-critical).
In the subfigures, nodes and edges that comprise the largest cluster on 
the lattice are colored black, all other nodes and edges are
colored grey.
\label{fig:2Dsamples}}
\end{figure}

Here, we consider the Euclidean relative neighborhood graph (RNG) for a planar
set of, say, $N$ points and determine the thresholds for bond and site percolation on 
this type of random graph. 
To prepone some of the details given in sect.\ \ref{sect:model}, note that a graph 
is subsequently referred to as $G=(V,E)$, where $V$ is a set of the nodes (represented 
by a set of $N$ distinct $d$-dimensional points, called nodes or sites, see Ref.\ \cite{Essam1970}), and where $E$ signifies the 
respective edgeset. Considering a particular metric, a certain ``length'' can further 
be associated with each edge (see discussion in sect.\ \ref{sect:model}).
In a RNG, $E$ contains proximity information regarding the close (spatial) neighbors for the terminal nodes of a given edge.
Here, bond percolation means that for a given instance of a RNG we 
occupy a fraction $p$ of the graph edges and assess the statistics
of clusters of adjacent sites connected by occupied edges. 
Examples of bond percolation for an instance of a RNG for different values of $p$ 
are shown in Fig.\ \ref{fig:2Dsamples}(a--c).

The RNG for a given set of points considering an Euclidean metric was 
introduced by Toussaint in 1980, see Ref.\ \cite{Toussaint1980}, 
who discussed its ability to extract perceptual relevant information from a planar set of points. 
This is relevant in the fields of computational geometry and pattern recognition, 
where important questions relate to the problem of finding structure behind the pattern
displayed by a set of points. 
RNGs find further application in the construction of planar ``virtual backbones'' for 
ad-hoc networks (i.e.\ collections of radio devices without fixed underlying
infrastructure), along which information can be efficiently transmitted \cite{Karp2000,Bose2001,Jennings2004,Yi2010}.
In Toussaints seminal article it was shown (by means of some illustrative examples) 
that, depending on the precise distribution of points in the plane, an instance of 
a RNG might behave similar to a minimum weight spanning tree (MST; i.e.\ a spanning tree
in which the sum of Euclidean edge weights is minimal, see Ref.\ \cite{clrs2001}) or a Delaunay 
triangulation (DT; in a DT, two nodes $i,j\in V$ are joined by an edge, if there is
a circle passing through them that contains no other points $k\in V\setminus \{i,j\}$, see Ref.\ \cite{CompGeom1985}) for the underlying set of points. 
Toussaint showed that for a planar set of points,
the MST is a (spanning) subgraph of the RNG, and further, the RNG is a (spanning) subgraph of the respective DT.
Finally, Ref.\ \cite{Toussaint1980} discusses two algorithms that allow to compute
the RNG for a given set of points, termed ALG-1 and ALG-2.
ALG-1 represents a naive implementation of the RNG construction rule (see sect.\ \ref{sect:model})
that terminates in time $O(N^3)$ and is correct for $d$-dimensional sets of points as well as for non-Euclidean metrics.
In contrast to this, ALG-2 is rather fast but limited to the planar case and to the Euclidean metric.
Being slightly more ``special'', ALG-2 is based on the observation that in the planar 
case and for an Euclidean metric the RNG is a subgraph of the DT. Since there 
are fast algorithms that allow to compute a DT for a planar set of points \cite{CompGeom1985,qHull} (terminating 
in time $O(N\log(N))$), 
a considerable speedup can be achieved, resulting in a worst case running time $O(N^2)$. 
Further, amending ALG-2 by standard techniques to accelerate ``range-queries'' 
yields an improved worst-case running time $O(N \log(N))$ (see discussion in sect.\ \ref{sect:model}).

As pointed out above, percolation on the DT of a given point set in the plane is 
well understood. The respective thresholds for site and bond percolation can, e.g.,
be found under Ref.\ \cite{wiki:PercThreshold}.
Regarding the subgraph hierarchy ${\rm MST} \subset {\rm RNG} \subset {\rm DT}$, the
question which subgraph of the DT still features a nontrivial percolation transition 
was addressed recently, see Ref.\ \cite{Billiot2010}. 
Intuitively, for the MST, the site and bond percolation thresholds 
are $1$, i.e.\ the transition points are trivial. 
Considering the RNG for a planar set of points and using the so called ``method of the 
rolling ball'', Ref.\ \cite{Billiot2010} established the existence of nontrivial 
site and bond percolation thresholds by analytic means. However, numerical estimates for
the transition points are not provided in the latter reference.
Here, to elaborate on that, we perform numerical simulations in order to determine the thresholds of
bond and site percolation on Euclidean RNGs for planar sets of points.
Since the RNGs are subgraphs of DTs we can expect that the critical exponents
that characterize the percolation transition on RNGs equal the
exponents of standard $2D$ percolation. The transition 
is hence expected to be in the $2D$ percolation universality class and
we are primarily interested in the site and bond percolation thresholds on 
Euclidean RNGs.

The remainder of the presented article is organized as follows.
In section \ref{sect:model}, we introduce RNGs and the algorithm we 
use in order to compute them in more detail.
In section \ref{sect:results}, we report on the numerical simulations and
the analysis of the topological and percolation properties of RNGs. 
Section \ref{sect:summary} concludes with a summary.

\section{Model and Algorithm\label{sect:model}}

\begin{figure}[t!]
\begin{center}
\includegraphics[width=1.0\linewidth]{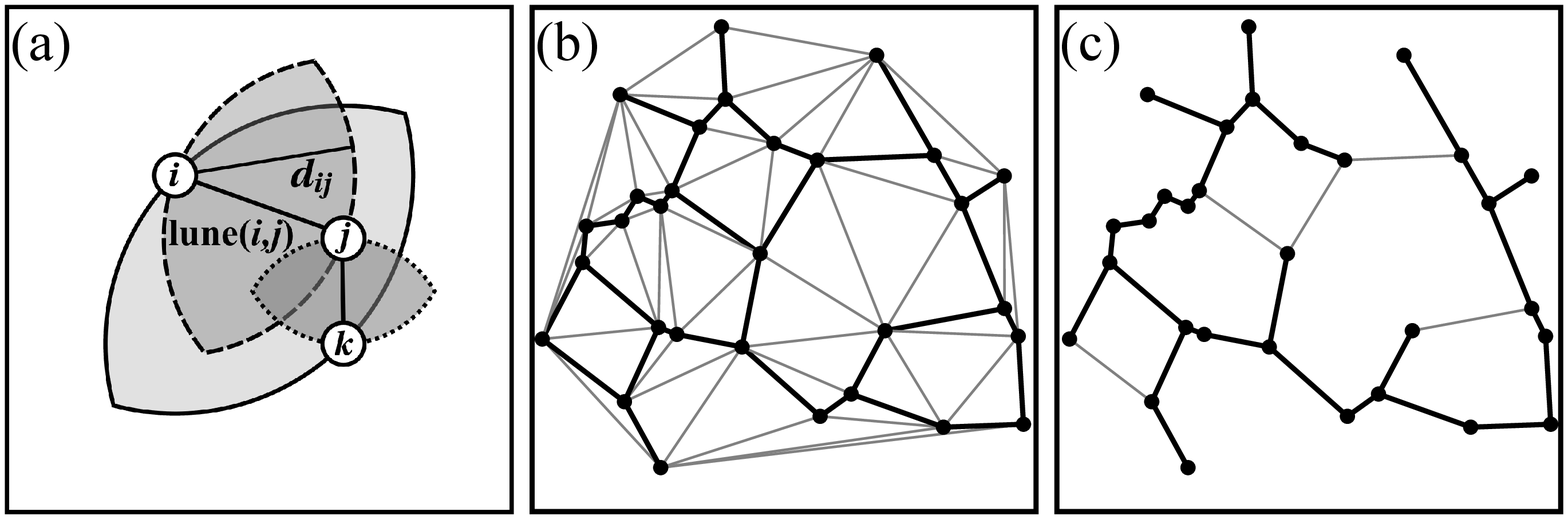}
\end{center}
\caption{Illustration of the Euclidean
relative neighborhood graph (RNG) and its relation to the Denlaunay triangulation (DT)
and minimum-weight spanning-tree (MST) for the same set of points. 
(a) In a RNG, two points, say $i\equiv p_i$ and $j\equiv p_j$, are connected by an undirected edge 
if no third point lies within ${\rm lune}(i,j)$ (see text for more details).
Further, a slightly less trivial example for $N=30$ points, showing
(b) the RNG (black edges) as spanning subgraph of the DT (black and grey edges), and,
(c) the MST (black edges) as spanning subgraph of the RNG (black and grey edges).
\label{fig:rngConstruction}}
\end{figure}

RNGs $G=(V,E)$ are based on the concept of \emph{relative closeness}. 
The nodeset of a $N$-point RNG is given by a set of $N$ distinct $d$-dimensional 
points, i.e.\ $V=\{p_1,p_2,\ldots,p_N\}$, where $p_i=(p_{i,1},\ldots,p_{i,d})$.
Further, consider a metric $L_r$ under which for two points 
$p_i$ and $p_j$ the distance measure $d_r(p_i,p_j)$ is given by 
\begin{align}
d_r(p_i,p_j)=\big[\sum\nolimits_{m=1}^d |p_{i,m}-p_{j,m}|^r\big]^{1/r}. \label{eq:metric}
\end{align}
Then, the edgeset $E$ of the RNG is obtained by connecting two points 
$p_i$ and $p_j$ using an undirected edge $\{p_i,p_j\}$ if they are at least as close 
to each other as to any third point $p_k$, see Fig.\ \ref{fig:rngConstruction}(a).
Hence, in order to get joined by an edge, the distance $d_{r}(p_i,p_j)$ of the two points has to satisfy the relation 
\begin{align}
d_{ij}\equiv d_r(p_i,p_j)\leq {\rm max}[d_r(p_i,p_k),d_r(p_j,p_k)] \label{eq:relDist}
\end{align}
for all $k=1\ldots N$, $k\neq i,j$.
If Eq.\ \ref{eq:relDist} is satisfied, then the two nodes are said to be \emph{relatively close}.
In geometrical terms, for each pair $p_i$ and $p_j$ of points, the respective 
distance $d_{ij}$ can be used to construct the \emph{lune} ${\rm lune}(p_i,p_j)$.
The lune is given by the intersection of two $d$-dimensional hyperspheres 
with equal radius $d_{ij}$ (with respect to the prevailing metric), 
which are centered at $p_i$ and $p_j$.
If no other point $p_k \in V\setminus\{p_i,p_j\}$ lies within ${\rm lune}(p_i,p_j)$, i.e.\ if the lune is empty,
Eq.\ \ref{eq:relDist} holds and $p_i$ and $p_j$ are thus relatively close.
In the remainder of the article, if not stated explicitly, we consider sets of points in dimension $d=2$ 
for the Euclidean metric $L_2$.

For a planar set of three points $(p_1,p_2,p_3)\equiv (i,j,k)$, the above ``selection criterion'' for 
RNG-edges is illustrated in Fig.\ \ref{fig:rngConstruction}(a).
Therein, the individual lunes are shown as shaded regions. 
Since ${\rm lune}(i,j)$ (encompassed by a dashed line) and ${\rm lune}(j,k)$ (encompassed by a dotted line) 
enclose no further point, the respective pairs of nodes are joined by undirected edges.
Only ${\rm lune}(i,k)$ (encompassed by a solid line) is not empty. It encloses the point $j$, hence 
the points $i$ and $k$ are not joined by an undirected edge. 
The resulting RNG is thus $G=(V=\{i,j,k\},E=\{\{i,j\},\{j,k\}\})$.
So as to facilitate intuition, a slightly less trivial example involving $N=30$ 
points is shown in Figs.\ \ref{fig:rngConstruction}(b-c).
In subfigure \ref{fig:rngConstruction}(b), the RNG is highlighted as subgraph of the DT, and,
in subfigure \ref{fig:rngConstruction}(c), the MST is indicated as subgraph of the RNG.

Note that a straight-forward implementation of the above RNG characteristics 
can be achieved by considering each pair of points (of which there are $O(N^2)$) and
checking whether one of the remaining $N-2$ points lies within their lune to rule out
that they are RNG neighbors. This, however,
yields an algorithm with running time $O(N^3)$, referred to as ALG-1 by Ref.\ \cite{Toussaint1980}
(however, note that a tremendous speed-up can be achieved by realizing that already a 
single point within a given lune is sufficient to rule out that the respective lune-defining
points are RNG-neighbors. Hence, as soon as for the lune of a particular pair of points the first such
``intruder'' is identified, one might safely proceed to the next pair of points).

\begin{figure}[t!]
\centerline{
\includegraphics[width=1.0\linewidth]{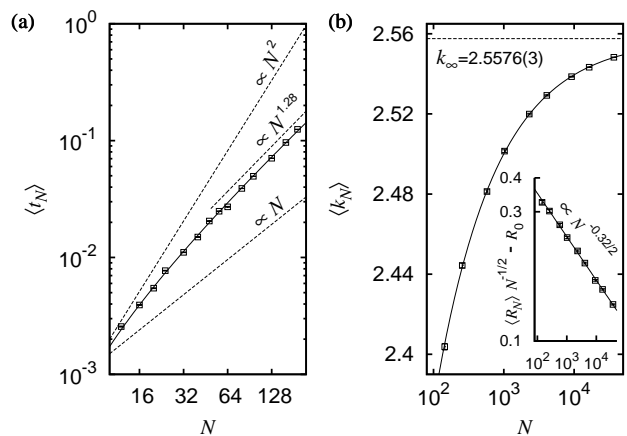}}
\caption{Finite-size scaling analysis for 
(a) the average running time $\langle t_N \rangle$ of the RNG algorithm for comparatively
small system sizes $N=12\ldots 192$,
(b) the main plot shows the average degree $\langle k_N\rangle$ of the RNG nodes, and the
inset illustrates the power-law correction to the scaling behavior of the average RNG diameter $\langle R_N\rangle$.
\label{fig:runTime_graphProp}} 
\end{figure}

For the planar case and for the Euclidean metric a more efficient algorithm, termed 
ALG-2 (see Ref.\ \cite{Toussaint1980}), can be devised. Based on the observation that
under the above assumptions the RNG is a subgraph of the DT, ALG-2 can be summarized by
the following two steps:
(i)  construct the DT $G_{\rm DT}=(V_{\rm DT},E_{\rm DT})$ for the planar set $V$ of points, and,
(ii) prune the edgeset $E_{\rm DT}$ of the DT by deleting all 
     $\{p_i,p_j\}\in E_{\rm DT}$ for which ${\rm lune}(p_i,p_j)$ is not empty.
The latter cleanup phase then results in the edgeset $E$ of the RNG for the underlying pointset.
So as to compute the DT of $V$ in step (i) above, we use the {\rm Qhull} computational
geometry library \cite{qHull} (the DT for a set of $N$ points can be computed in 
time $O(N\log(N))$ \cite{CompGeom1985,qHull}).
For the implementation of step (ii) we use the ``cell-list'' method. Therein, the unit square, within 
which the $N$ points are distributed uniformly at random, is subdivided into $L \times L$ cells (where $L=\sqrt{N}$), 
and the $2D$ cell-indices $(i_1,i_2)=(\lfloor L\cdot p_{i,1} \rfloor,\lfloor L\cdot p_{i,2}\rfloor)$ for all points $p_i \in V$ are
determined. Each cell then is equipped with a list of the points it contains. 
If the lune of a particular pair $\{p_i,p_j\}$ of points then needs to be checked for emptiness, 
only a small number $n_{ij}$ of cells close to the cells with indices $(i_1,i_2)$ and $(j_1,j_2)$ have to be 
addressed to reach all candidate points. Note that the number $n_{ij}$ of cells 
to be checked depends on the precise distance $d_{ij}$ between the respective points. 
Typically, for large $N$, $n_{ij}$ is small for points located in the ``bulk'' of 
the unit square, and $n_{ij}$ can be rather large for points that are located along the circumference of the convex hull of
$V$. The speed-up achieved by the cell-method is quite impressive, yielding an improved 
algorithm ALG-2-CELL that terminates in time $O(N \log(N))$. 
This is illustrated in Fig.\ \ref{fig:runTime_graphProp}(a), where for small systems of size $N<200$
the average running time $\langle t_N \rangle$ (averaged over $500$ $N$-point instances) is shown.
The solid line indicated in the figure is a fit to a function of the form $\langle t_N\rangle\propto N \log(N)$.
However, note that for $N$ not too small the data also fits well to an effective power-law function that 
increases $\propto N^{1.28(2)}$.

Subsequently, we will use ALG-2-CELL to compute the Euclidean RNG for $2D$ sets of points and we 
compute the bond and site percolation thresholds on these graphs.


\section{Results \label{sect:results}}

In the current section we will use ALG-2-CELL to compute the Euclidean RNG for planar sets of 
$N=144 (=12^2) \ldots 36864 (=192^2)$ points, where results are averaged over $2000$ independent
RNG instances.
In subsect.\ \ref{subsect:res_top}, we report on some topological properties of RNGs and then we 
compute the bond and site percolation thresholds on these graphs.
The data analysis for the respective bond percolation problem is discussed in detail
in subsect.\ \ref{subsect:res_bondPerc},
and the discussion for the site percolation problem in subsect.\ \ref{subsect:res_sitePerc} is kept more brief.
A visual account of bond percolation on an instance of a $N=512$ RNG, i.e.\ a $2D$ system of effectively $L\times L$ points, 
is given in Fig.\ \ref{fig:2Dsamples}(a-c).

\subsection{Topological properties of planar RNGs}
\label{subsect:res_top}

First, we discuss two topological characteristics of RNGs, namely the average node degree
and diameter (i.e.\ the longest among all shortest paths) of the RNG in the 
limit $N\to \infty$. Bear in mind that the nodes of a $N$-point 
RNG are distributed uniformly at random in the unit square. 
Thus, one might expect that the scaling behavior of observables in a $2D$ RNG depends
on the effective system length $L=N^{1/2}$.

\paragraph{Average degree of the RNG:}
The scaling behavior of the average degree $\langle k_N \rangle$ as function of the RNG size
$N$ is shown in Fig.\ \ref{fig:runTime_graphProp}(b). 
Therein, the solid line indicates a fit to the function $\langle k_N \rangle= k_\infty - a N^{-b}$
where we find the asymptotic average degree $k_\infty=2.5576(3)$, $a=1.85(5)$, and $b=0.503(5)$ 
for a reduced chi-square value $\chi^2_{\rm red}=0.76$ (considering $n_{\rm dof}=6$ degrees of freedom).

\begin{table}[b!]
\caption{\label{tab:tab1}
Critical properties that characterize bond and site percolation
(BP and SP, respectively) on Euclidean RNGs for planar sets of points. 
From left to right: 
Critical point $p_c$ (obtained from the analysis of the Binder ratio), critical exponents $\nu$ and $\beta$ obtained 
from the order parameter, and $\gamma$, obtained from the
order parameter fluctuations and the scaling 
behavior of the average size of the finite clusters for BP and SP, respectively.
} 
\begin{ruledtabular}
\begin{tabular}[c]{l@{\quad}llllll}
  Type & $p_c$  & $\nu$ & $\beta$ & $\gamma$ \\
\hline
RNG-BP  & 0.771(2)   & 1.33(6)  & 0.15(2) & 2.40(6)  \\
RNG-SP  & 0.796(2)   & 1.33(6) & 0.14(3) & 2.39(7) \\
\end{tabular}
\end{ruledtabular}
\end{table}

\paragraph{Average diameter of the RNG:}
The average diameter $\langle R_N \rangle$ of the RNG as function of $N$ is summarized in 
the inset of Fig.\ \ref{fig:runTime_graphProp}(b). For a planar graph like the RNG one 
can already expect the approximate scaling behavior $\langle R_N \rangle\propto N^{1/2}$.
Upon analysis we found that the data fits best to a function of the form
$\langle R_N \rangle = R_0 N^{1/2}[1+bN^{-\omega/2}]$, where $\omega$ indicates
a correction-to-scaling exponent. We estimate $R_0=1.75(2)$, $b=0.42(1)$, and $\omega=0.32(3)$
for a reduced chi-square value $\chi^2_{\rm red}=1.64$ (considering $n_{\rm dof}=6$ degrees of freedom).
In the inset of Fig.\ \ref{fig:runTime_graphProp}(b) we aimed to extract the correction to scaling according to
$\langle R_N \rangle N^{-1/2} - R_0\propto N^{-\omega/2}$ .
The numerical value of $R_0$ can also be set into a context \cite{Toussaint1980}: for the MST on has 
$\langle R_N^{\rm MST} \rangle=R_0^{\rm MST} N^{1/2}$, where $0.5\leq R_0^{\rm MST}\leq 0.707$, see Refs.\ \cite{Gilbert1965,Roberts1968}.
Since the RNG is a supergraph of the MST one thus might expect to find $R_0 \geq R_0^{\rm MST}$.

\begin{figure}[t!]
\centerline{
\includegraphics[width=1.0\linewidth]{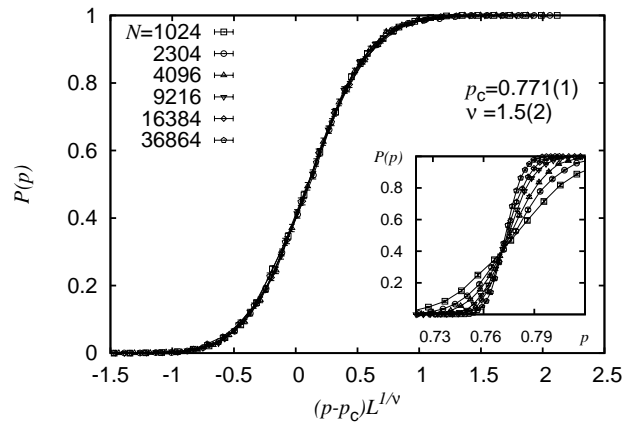}}
\caption{Finite-size scaling analysis of the bond percolation probability $P(p)$ 
 (i.e.\ the probability of simultaneous percolation along both independent directions; see text) 
on RNGs for planar sets of $N=1024\ldots 36864$ points, 
averaged over $2000$ different graph instances.
The main plot shows the data collapse obtained using 
Eq.\ \ref{eq:scalingAssumption}, and the inset illustrates
the raw data close to the critical point.
\label{fig:percProb_fss}} 
\end{figure}  
\subsection{Results for bond percolation on planar RNGs}
\label{subsect:res_bondPerc}

To simulate the bond-percolation problem on instances of RNGs we implemented the highly efficient, 
union-find based algorithm due to Newman and Ziff, see Ref. \cite{Ziff2000,Ziff2001}.
Therein, initially, each node comprises its own (single-node) cluster. 
We proceed by choosing edges from $E$, one after the other, in random order. 
For each edge we check whether its terminal nodes belong to different clusters. If this is 
the case, the respective clusters are merged using the ``union-by-size'' approach. 
Once all edges are considered, the particular RNG instance is completed.
In principle, this allows to compute observables very efficiently with a resolution
of $O(1/|E|)$. For a more clear presentation, and so as to be able to 
compute proper errorbars for the observables below, we consider $2000$ independent
RNG instances for a given value of $N$ and approximately $80$ supporting points on 
the $p$ axis (in the vicinity of the critical point) for which averages are computed. 
The observables we consider below can be rescaled following a common scaling assumption.
Below, this is formulated for a general observable $y(p,L)$.
This scaling assumption states that if the observable obeys scaling, it can be rewritten as
\begin{eqnarray}
y(p,L)= L^{-b}~f[(p-p_c) L^{1/\nu}], \label{eq:scalingAssumption}
\end{eqnarray}
wherein $\nu$ and $b$ represent dimensionless critical exponents (or ratios thereof, see below),
$p_c$ signifies the critical point, and $f[\cdot]$ denotes an unknown scaling function.   
Following Eq.\ \ref{eq:scalingAssumption}, data curves of the observable $y(p,L)$ recorded at 
different values of $p$ and $L$ \emph{collapse}, i.e.\ fall on top of each other, if $y(p,L) L^{b}$ 
is plotted against the combined quantity $\epsilon \equiv (p-p_c) L^{\nu}$ and if further 
the scaling parameters $p_c$, $\nu$ and $b$ that enter Eq.\ \ref{eq:scalingAssumption} are chosen properly.    
The values of the scaling parameters that yield the best data collapse determine the numerical 
values of the critical exponents that govern the scaling behavior of the underlying observable $y(p,L)$.
In order to obtain a data collapse for a given set of data curves we here perform a
computer assisted scaling analysis, see Refs.\ \cite{houdayer2004,autoScale2009}.
The resulting numerical estimates of the critical thresholds and exponents for bond and site percolation on 
planar Euclidean RNGs are listed in Tab.\ \ref{tab:tab1}. 
In the subsequent paragraphs, we report on the results found for different observables.

\paragraph{Percolation probability:}
In order to provide a measure of percolation probability for the $2D$ RNG, we proceed as follows:
For each instance of a $N$-point RNG we first determine the $L$ points that are closest
to the left, right, top, and bottom boundary. As a sufficient condition for percolation along, say,
the horizontal direction, we consider the event that a point on the left and the right boundary are
part of the same cluster. Here, we put under scrutiny the particular event that the system simultaneously 
percolates along both independent directions (other criteria yield similar results).
The finite-size scaling analysis of the corresponding percolation or ``spanning'' probability $P(p)$ is
summarized in Fig.\ \ref{fig:percProb_fss}. Setting $b=0$ in Eq.\ \ref{eq:scalingAssumption} 
(as appropriate for a dimensionless quantity) and restricting the analysis to the interval 
$\epsilon=[-0.5,0.5]$ on the rescaled $p$-axis, we find that $p_c=0.771(1)$ and $\nu=1.5(2)$ yield a best data collapse
with ``collapse-quality'' $S=0.33$ (the numerical value of $S$ measures the mean--square distance of the data points to the master
curve, described by the scaling function, in units of the standard error
\cite{houdayer2004}).
Note that the numerical value of the correlation length
exponent is in agreement with the standard $2D$ percolation exponent $\nu=4/3\approx 1.333$.

\begin{figure}[t!]
\begin{center}
\includegraphics[width=1.0\linewidth]{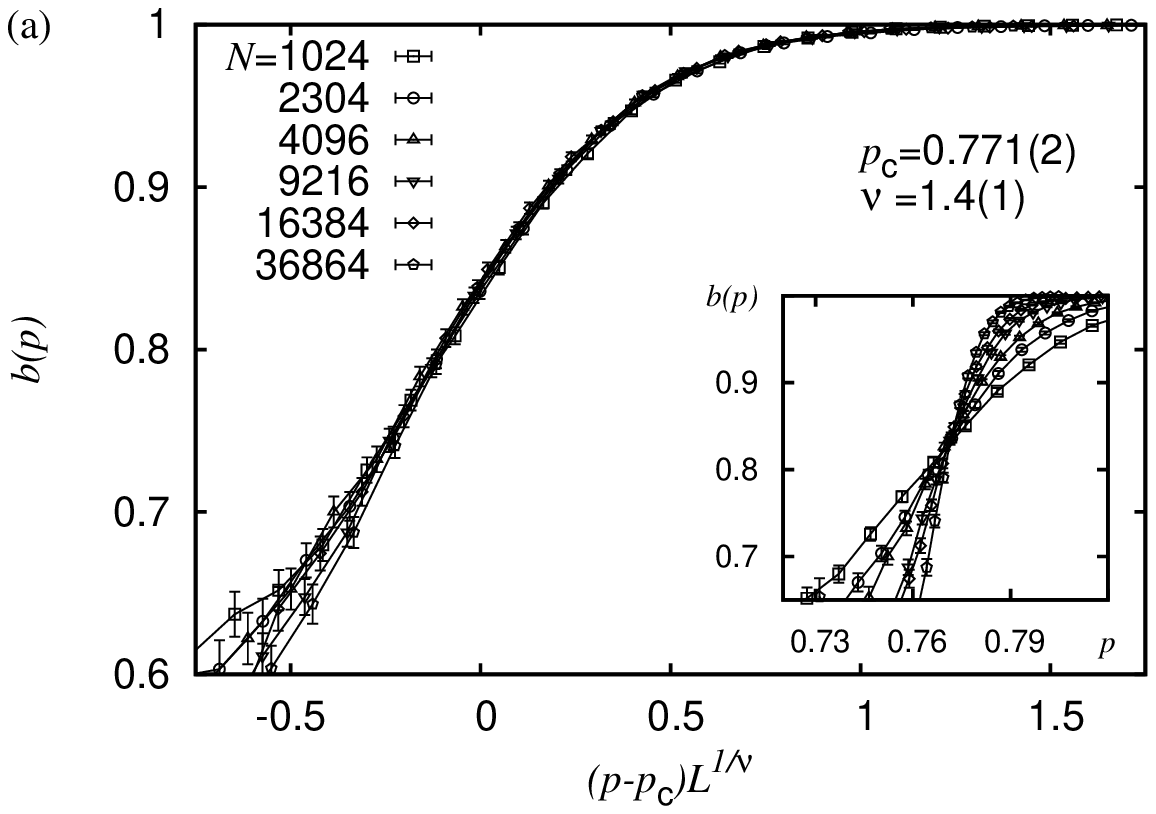}
\includegraphics[width=1.0\linewidth]{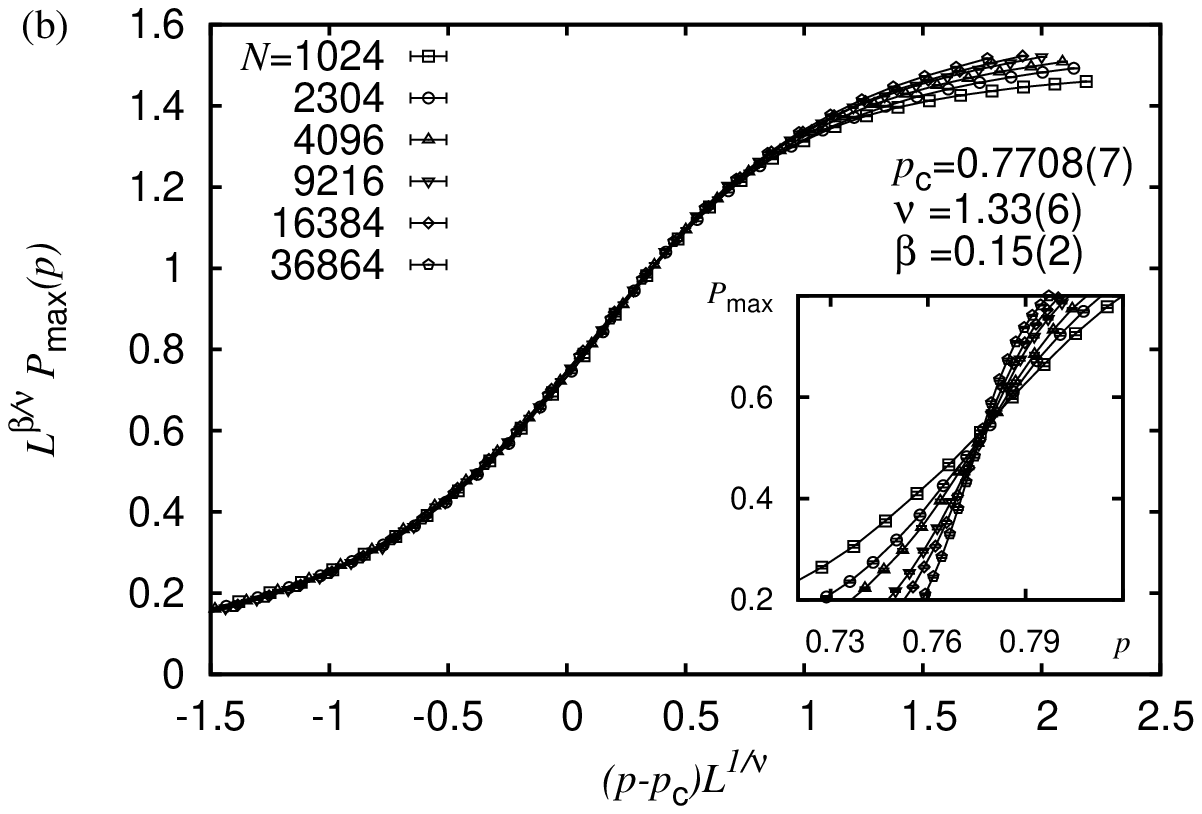}
\includegraphics[width=1.0\linewidth]{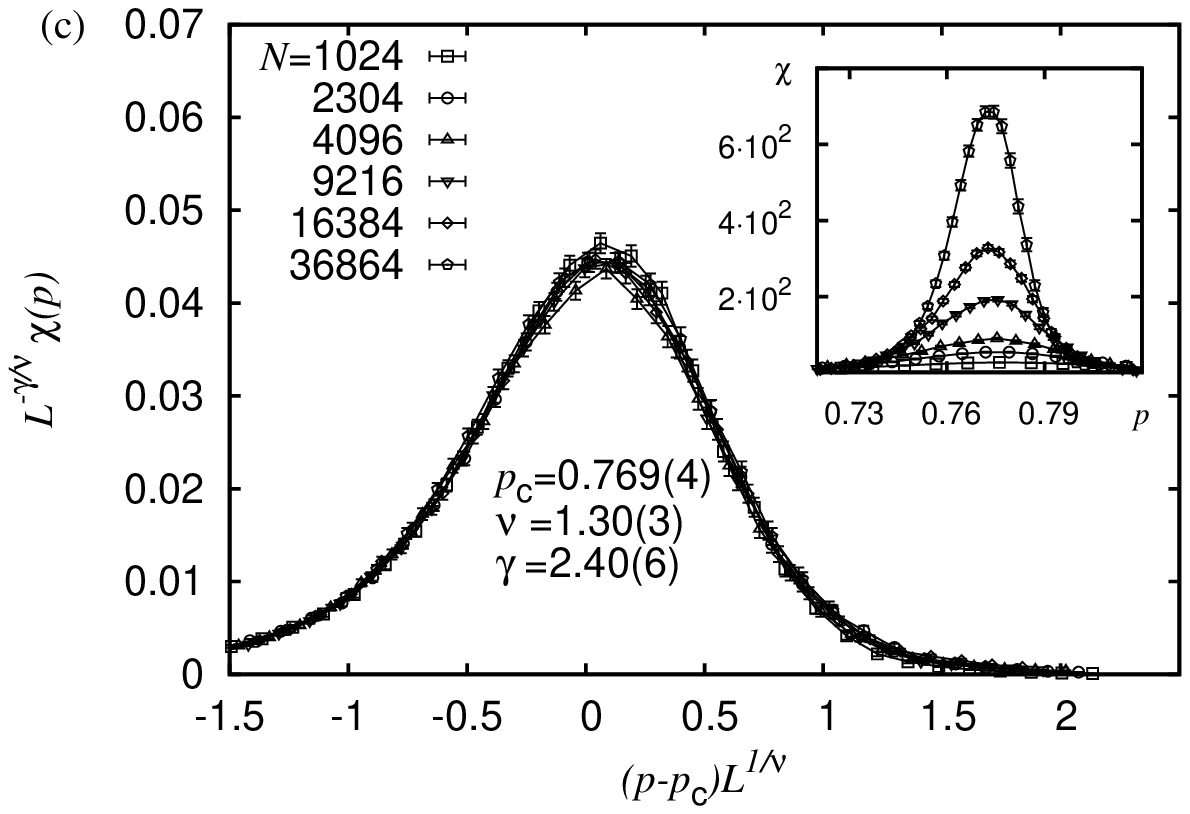}
\end{center}
\caption{Finite-size scaling analyses related to the 
relative size $s_{\rm max}$ of the largest cluster 
of sites on RNGs for planar sets of $N=1024\ldots 36864$ points, 
averaged over $2000$ graph instances. The main plots show the data collapse obtained according
to Eq.\ \ref{eq:scalingAssumption}, and the insets illustrate
the raw data close to the critical point.
The subfigures show different ways to analyze $s_{\rm max}$
in terms of
(a) the Binder ratio $b(p)$,
(b) the order parameter $s_{\rm max}(p)$, and,
(c) the fluctuation $\chi(p)=N{\rm var}(s_{\rm max})$ 
of the order parameter.
\label{fig:orderPar_fss}}
\end{figure}  
\paragraph{Order parameter statistics:}

As a second observable we consider $s_{\rm max}$, i.e.\ the relative size 
of the largest cluster of points joined by edges.
In this regard, a further dimensionless quantity commonly referred to in the analysis of phase
transitions is the \emph{Binder ratio} \cite{binder1981}
\begin{eqnarray}
b(p) = \frac{1}{2} \Big[3 - \frac{\langle s_{\rm max}^4(p) \rangle}{ \langle s_{\rm max}^2(p) \rangle^2} \Big].   \label{eq:binderPar}
\end{eqnarray}
This ratio of moments scales according to Eq.\ (\ref{eq:scalingAssumption}), where, as for the spanning probability above, $b=0$.
As can be seen from the inset of Fig.\ \ref{fig:orderPar_fss}(a), the Binder ratio exhibits 
a nice common crossing point of the data curves for different values of $N$.
The best data collapse (obtained in the (unsymmetrical) range $\epsilon \in [-0.1,1.0]$)
yields $p_c=0.772(2)$, and $\nu=1.4(2)$ with a quality $S=1.22$.
As evident from the rescaled data (main plot of Fig.\ \ref{fig:orderPar_fss}(a)),
there are rather strong deviations from the expected scaling behavior as $p<p_c$.
To account for this, the scaling analysis is performed in the rather unsymmetrical
interval $\epsilon \in [-0.1,1.0]$ on the rescaled $p$-axis to accentuate the 
region $p>p_c$ where $b(p)$ seems to scale well.
As above, the numerical value of the exponent $\nu$ is in agreement with the 
standard $2D$ percolation exponent.
Considering the \emph{order parameter} 
\begin{eqnarray}
P_{\rm max}(p)=\langle s_{\rm max}(p) \rangle, \label{eq:orderPar}
\end{eqnarray}
the best data collapse (obtained in the range $\epsilon \in [-1,1]$)
yields $p_c=0.7708(7)$, $\nu=1.33(6)$, and $\beta=0.15(2)$ with a quality $S=0.50$, see Fig.\ \ref{fig:orderPar_fss}(b)).
If we fix the numerical values of the critical exponents to the expected values $\nu=4/3\approx 1.333$ and $\beta=5/36\approx0.139$ 
we are left with only one adjustable parameter, resulting in the estimate
$p_c=0.7711(6)$ with $S=0.88$.
A further critical exponent can be estimated from the scaling of the \emph{order parameter fluctuations} $\chi(p)$,  given
by
\begin{eqnarray}
\chi(p)= N [ \langle s_{\rm max}^2(p) \rangle - \langle s_{\rm max}(p) \rangle^2  ]. \label{eq:suscept}
\end{eqnarray}
A best data collapse for this observable (attained in the range $\epsilon \in [-0.2,0.2]$) 
results in the estimates $p_c=0.769(4)$, $\nu=1.30(3)$, and $\gamma=2.40(6)$ with a quality $S=0.24$, see Fig.\ \ref{fig:orderPar_fss}(c).  
Note that the numerical value of the fluctuation exponent is in agreement with the 
expected value $\gamma=43/18\approx 2.389$. 

\paragraph{Average size of the finite clusters:}
As a last observable we consider the average size $\langle S_{\rm fin} (p)\rangle$ of all finite clusters for a particular 
RNG, averaged over different instances of RNGs.
The respective definition reads \cite{Sur1976,stauffer1994}
\begin{eqnarray}
S_{\rm fin}(p) = \frac{\sum_{s}^\prime s^2\, n_s(p)}{\sum_{s}^\prime s\, n_s(p)}, \label{eq:finClusters}
\end{eqnarray}
where $n_s(p)$ signifies the probability mass function of cluster sizes for a single RNG instance 
at a given value of $p$. Note that the sums run over finite clusters only \cite{Sur1976,stauffer1994} 
(indicated by the prime), i.e.\ if the precise  
configuration features a system spanning cluster (spanning horizontally or vertically or both), 
this cluster is excluded from the sums that enter Eq.\ \ref{eq:finClusters}.
The average size of all finite clusters is expected to scale according to Eq.\ \ref{eq:scalingAssumption},
where $b=-\gamma/\nu$.
Restricting the data analysis to the interval $\epsilon\in [-1.0,1.0]$ on the rescaled $p$-axis, the 
optimal scaling parameters are found to be $p_c=0.770(2)$, $\nu=1.36(4)$, and $\gamma=2.33(5)$
with a quality $S=0.61$ (not shown). 
Note that here, the numerical values of the extracted exponents are in reasonable agreement with 
the expected values and the estimate of the critical threshold for bond percolation is consistent 
with the numerical values found above. 

\subsection{Results for site percolation on planar RNGs}
\label{subsect:res_sitePerc}

The analysis in terms of the site percolation problem was carried out analogous to that
of the bond percolation problem in the preceding subsection.
However, we here list only the estimates of the critical points obtained from 
the data collapse analysis for the different observables. In this regard 
we have $p_c=0.799(1)$ (percolation probability), $p_c=0.796(2)$ (Binder ratio), 
$p_c=0.795(1)$ (order parameter), $p_c=0.794(4)$ (order parameter fluctuation; see Tab.\ \ref{tab:tab1}), 
$p_c=0.798(3)$ (average size of the finite clusters).
Further, the critical exponents $\nu=1.33(6)$ and $\beta=0.14(3)$ obtained from the order parameter 
and $\gamma=2.39(7)$, obtained from the scaling behavior of the average size of the finite clusters,
are listed in Tab.\ \ref{tab:tab1}.

\section{Discussion and summary \label{sect:summary}}
In the presented article we have closely investigated the statistical 
and percolation properties of planar Euclidean RNGs via numerical simulations.
In regard of the subgraph hierarchy $MST\subset RNG \subset DT$,
recently the question was raised which subgraph of the DT (on which
the percolation problem is well studied) still features a nontrivial
percolation transition \cite{Billiot2010}. 
Intuitively, for the MST this is not the case. For the RNG previous 
analytic studies established the existence of nontrivial site and 
bond percolation thresholds \cite{Billiot2010}, but no numerical estimates where 
provided.
Here we quote $p_{c,{\rm bond}}^{\rm RNG}=0.771(2)$ and $p_{c,{\rm site}}^{\rm RNG}=0.796(2)$,
obtained by means of finite-size scaling analyses in terms of the ``data-collapse'' technique.
Further, we also deduced the critical exponents that govern both percolation 
transitions on the RNG and found them to be consistent with those that describe
the standard $2D$ percolation phenomenon (as expected).
So as to yield maximally justifiable results through numerical redundancy, 
we considered various observables to estimate the critical points and exponents.

As discussed in subsect.\ \ref{subsect:res_top}, the asymptotic average degree
for the RNG reads $k_\infty=2.5576(3)$. In order to put the above critical points 
into a context, we might attempt to compare them to the threshold values for 
regular $2D$ lattices with a similar degree. 
E.g., the site and bond percolation thresholds for 
the ($3$,$12^2$)-Archimedean lattice, having degree $k=3$,
read $p_{c,{\rm bond}}=0.74042081(10)$ \cite{Ziff2009} and $p_{c,{\rm site}}=0.807900764\ldots$ \cite{Suding1999}.
Both are actually not thus far from the respective thresholds on the RNG.
(Further, for the $k=3$ Martini lattice one has 
$p_{c,{\rm bond}}=0.707107\ldots$ \cite{Ziff2006} and $p_{c,{\rm site}}=0.764826\ldots$ \cite{Scullard2006}.)
Regarding $2D$ random lattices, site and bond percolation on the Voronoi tesselation 
of a planar pointset, also having degree $k=3$, give rise to the threshold values
$p_{c,{\rm bond}}=0.666931(5)$ and $p_{c,{\rm site}}=0.71410(2)$ \cite{Becker2009}.
In addition, site percolation on planar $\Phi^3$ random graphs result in $p_{c,{\rm site}}=0.7360(5)$ \cite{Kownacki2008}.
In comparison, the estimates from the latter random graphs
are less close to the estimates for the RNG as compared to the ($3$,$12^2$)-Archimedean lattice thresholds.

As pointed out in the introduction, RNGs are discussed in the context of the construction of 
planar ``virtual backbones'' for ad-hoc networks that guarantee connectedness of all considered 
nodes \cite{Karp2000,Bose2001,Jennings2004,Yi2010}.
In this regard, from a point of view of stability, it would be interesting to 
quantify how susceptible RNGs are with respect to a random failure and targeted 
removal of nodes and to compare the results to other ``proximity graph''
instances 
which are discussed in the same context. 
Such investigations are currently underway \cite{furtherStudies}.


\begin{acknowledgments}
I am much obliged to A.~K.\ Hartmann and
C.\ Norrenbrock for valuable discussions and for comments that
helped to amend the manuscript.
Further, I gratefully acknowledge financial support from the DFG 
(\emph{Deutsche Forschungsgemeinschaft}) under grant HA3169/3-1.
The simulations were performed at the HPC Cluster HERO, located at 
the University of Oldenburg (Germany) and funded by the DFG through
its Major Instrumentation Programme (INST 184/108-1 FUGG) and the
Ministry of Science and Culture (MWK) of the Lower Saxony State.
\end{acknowledgments}


\bibliography{rng_perc.bib}

\end{document}